\documentclass[preprint,showkeys,preprintnumbers,amsmath,amssymb]{revtex4}

\usepackage{graphicx}
\usepackage{dcolumn}
\usepackage{bm}

\def\e{\mathrm{e}}

\begin{document}

\author{R. Rossi Jr.}
\email{romeu.rossi@ufv.br}
\affiliation{Universidade Federal de Vi{\c c}osa - Campus Florestal,
LMG818 Km6, Minas Gerais, Florestal 35690-000, Brazil}

\title{Restrictions for the Causal Inferences in an Interferometric System}

\begin{abstract}
Causal discovery algorithms allow for the inference of causal structures from probabilistic relations of random variables. A natural field for the application of this tool is quantum mechanics, where a long-standing debate about the role of causality in the theory has flourished since its early days. In this paper, a causal discovery algorithm is applied in the search for causal models to describe a quantum version of Wheeler's delayed-choice experiment. The outputs explicitly show the restrictions for the introduction of classical concepts in this system. The exclusion of models with two hidden variables is one of them. A consequence of such a constraint is the impossibility to construct a causal model that avoids superluminal causation and assumes an objective view of the wave and particle properties simultaneously. 
\end{abstract}

\maketitle

The search for stable connections that explain the correlation between two events is the main goal of natural science. A simple indication, based on statistical results, that event B will follow event A is not enough to fulfill one's scientific expectations. Causal relations are required to assure the persistence of the conjunction A-B, and their robustness allows us to construct theories with predictive power.

Recently, a formal approach for causal relations, developed as a result of contributions from different fields (statistics, philosophy, and computer science) was summarized in \cite{pearl, spirtes}. The mathematical formalism of the theory promoted a revolutionary development in investigations about counterfactuals, interventions, and the inferences of causal models from statistical data. In this theory, causal structure is represented by a Directed Acyclic Graph (DAG), which is composed of nodes and directed edges connecting those nodes. The acyclicity certifies that no paths in the graph begin and end at the same node. The nodes represent variables of the system to be model and the directed edges are associated with the causal connections between those variables. This formalism has attracted much attention from different areas \cite{dif1,dif2,dif3}. The role of causality in quantum systems has been a theme of discussions since the early days of quantum theory. Hence, the application of the new formalism of causality in the context of quantum mechanics is of great interest \cite{art6,art7,art8,art9,art10,art11,art111,art112,art12,art13}.

In this approach of causality, strategies to infer causal models from statistical data are known as causal discovery algorithms (a brief summary is given in the appendix). In Ref. \cite{art13}, the authors investigate the possibility of inferring a causal structure for the correlations that have emerged from an entangled system. They show that the algorithm can not distinguish correlations that violate Bell inequalities from correlations that satisfy them. They also show that the possible candidates of causal structures that would describe the correlations of the entangled system have to infringe one of the basic principles of these algorithms; the stability assumption, which states that ``an observed statistical independence between variables should not be explained by fine-tuning of the causal parameters''\cite{art13}.

When applied to quantum systems, causal discovery algorithms allow us to explicitly show the restrictions on inferences of causal relations. Such inferences are usually motivated by our classical intuitions, and the debate about causal inferences involving hidden variables is usually present in the literature of Bell inequality. In \cite{Ionicioiu2}, the authors conduct the discussion about causality and hidden variables in a different system; a quantum version of Wheeler's delayed-choice experiment. In this context, the hidden variables are considered as a possible cause of the wave-like or particle-like behavior of photons in the experiment. The authors show that, in this system, an objective view of the wave and particle properties using hidden variables is not consistent. This setup was implemented experimentally in \cite{art1,art2,art3,art4,art5}.

In Ref.\cite{Ionicioiu1}, the authors present an extension of the system studied in \cite{Ionicioiu2}. The system in \cite{Ionicioiu1} is composed of a photon that goes through a Mach-Zehnder interferometer (MZI) with a removable second beamsplitter (BS2) controlled by a quantum device described by a qubit $q_{b}$. The states of a qubit $q_{a}$ correspond to the two exits of the interferometer. A third subsystem is included; the qubit $q_{c}$ prepared in an entangled state with $q_{b}$. With the inclusion of the entanglement between $q_{b}$ and $q_{c}$, the authors considered a second hidden variable, independent of the first one, that allows for an analysis that shows the incompatibility between determinism, wave–particle objectivity, and local independence in the interferometric system.

In the present paper a causal discovery algorithm is applied to the system presented in Ref.\cite{Ionicioiu1}. The algorithm explicitly reveals the restrictions over the possible causal assumptions that can be inferred for this system. As an output, the algorithm yields groups of possible causal structures that depend on the detection of the ordering of the qubits ($q_{a}$, $q_{b}$ and $q_{c}$). The analysis of these groups shows the restrictions regarding the causal assumptions. One of the restriction is the impossibility of constructing a causal structure with two hidden variables for this system, and this precludes the existence of a causal model that avoids superluminal causation and assumes an objective view of the wave and particle properties simultaneously.

The causal discovery algorithm allows us to critically analyze the theory of hidden variables. In Ref.\cite{Ionicioiu1}, contradictions between the predictions of quantum mechanics and of a hidden variable theory are also shown. However, the strategies used and the main results are different from the ones reported here. In Ref.\cite{Ionicioiu1}, the authors show that the quantum mechanical probability distribution cannot result from the probability distribution given by the hidden variable theory that assumes wave–particle objectivity, local independence, and determinism. Here, the causal structures that conform with the correlations given in the interferometric system are investigated through the application of a causal discovery algorithm. Some of these causal structures, given as outputs of the algorithm, have hidden variables. Therefore, the analysis of these outputs gives us the information about the role played by hidden variables i.e., to which variables can they be connected to, how many of them can be used in the model, etc.

The causal structures obtained here, from the algorithm are in accordance with the negative result deduced in \cite{Ionicioiu1}. However, no reference to determinism is necessary here. As it will be shown, the causal structures given by the algorithm do not allow the presence of two hidden variables. The assumption of local independence and wave–particle objectivity considered in \cite{Ionicioiu1}, demanded the existence of one hidden variable associated with the photon ($q_{a}$) and a second one associated with the subsystem $q_{b}-q_{c}$. Therefore, a consequence of the algorithm applied in the interferometric system is that no causal structure can simultaneously fulfill the conditions of local independence and wave–particle objectivity, and no deterministic assumption is needed to obtain this negative result. This comparison shows how causal discovery algorithm allows for new insights in the analysis of the theories on hidden variables in quantum mechanics.

For a causal discovery algorithm, the input is a set of conditional independence relations held among the variables of interest. We define the variables $A$, $B$, and $C$, corresponding to the measurement results of $q_{a}$, $q_{b}$, and $q_{c}$, respectively. The states of $q_{a}$: $|0\rangle_{A}$ and $|1\rangle_{A}$ represent the exits of the interfrometer. $q_{b}$: $|0\rangle_{B}$ and $|1\rangle_{B}$ describe the absence and presence of the second beamsplitter, respectively. In the experimental implementation proposed in \cite{Ionicioiu1}, $q_{b}$ and $q_{c}$ are also photons and $|0\rangle$ ($|1\rangle$) represents absence (presence) of a photon in the channel $B$ or $C$. Fig.1 shows the quantum circuit diagrams for this setup. The qubit $q_{c}$ can be rotated before detection in $D_{C}$.

\begin{figure}[h]
\centering
  \includegraphics[scale=0.3]{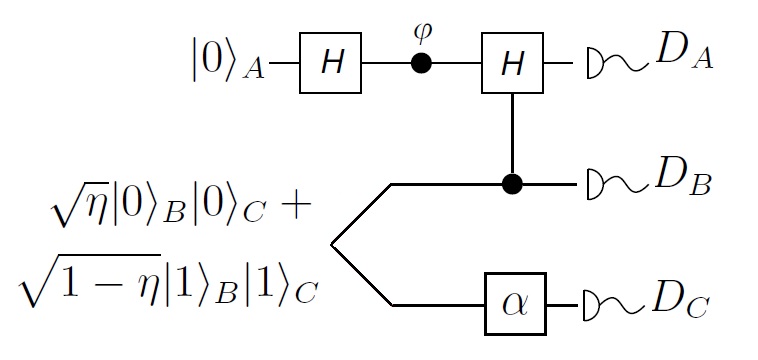}\\
  \caption{Quantum circuit diagrams for the setup proposed in Ref.\cite{Ionicioiu1}.}
\end{figure}

The initial state of the global system is given by:

\begin{equation}
|\psi_{0}\rangle = |0\rangle_{A}\left(\sqrt{\eta}|0\rangle_{B}|0\rangle_{C} + \sqrt{1-\eta}|1\rangle_{B}|1\rangle_{C}\right),
\end{equation}
where $\eta$ is a parameter that regulates the entanglement between $q_{b}$ and $q_{c}$.

The state just before the measurements can be written as:

\begin{eqnarray}
|\psi\rangle &=& \left(\sqrt{\eta}\cos(\alpha)|p\rangle_{A}|0\rangle_{B} + \sqrt{1-\eta}\sin(\alpha)|\omega\rangle_{A}|1\rangle_{B}\right)|0\rangle_{C}\notag\\
&-&\left(\sqrt{\eta}\sin(\alpha)|p\rangle_{A}|0\rangle_{B} + \sqrt{1-\eta}\cos(\alpha)|\omega\rangle_{A}|1\rangle_{B}\right)|1\rangle_{C}\label{vstate} ,
\end{eqnarray}
where $\alpha$ is a parameter related to the rotation that $q_{c}$ undergoes before the measurement. The particle-like state and the wave-like state are given by $|p\rangle=\frac{1}{\sqrt{2}}(|0\rangle + e^{i\phi}|1\rangle)$ and $|\omega\rangle=\e^{i\phi / 2 }(\cos(\phi/2)|0\rangle - i\sin(\phi/2)|1\rangle)$ respectively.

Two variables $X$ and $Y$ are considered conditionally independent given $Z$ if one of the equivalent condition is fulfilled:

\begin{eqnarray}
P(X|Y,Z)&=&P(X|Z)\notag\\
P(Y|X,Z)&=&P(Y|Z)\notag\\ 
P(X,Y|Z)&=&P(X|Z)P(Y|Z) \label{cond}
\end{eqnarray}   

We denote the conditional independence between $X$ and $Y$, given $Z$ as $X\perp\!\!\!\perp Y | Z$.

From the conditions given in eq.(\ref{cond}), where the probabilities can be calculated from the state in eq.(\ref{vstate}), one can obtain the set of conditional independence relations for the qubits $q_{a}$, $q_{b}$, and $q_{c}$. The results of this calculation are the relations: $A\perp\!\!\!\perp C | B$ and $C\perp\!\!\!\perp A | B$ (the second can be inferred from the first using the semi-graphoid axioms \cite{pearl}).

Using the relation $A\perp\!\!\!\perp C | B$ as an input for the algorithm IC* presented in \cite{pearl}, the output returned is the pattern shown in Fig.2, where the undirected edge with a circle at both extremities represents any of the five possibilities shown in Fig. 3.

\begin{figure}[h]
\centering
  \includegraphics[scale=0.08]{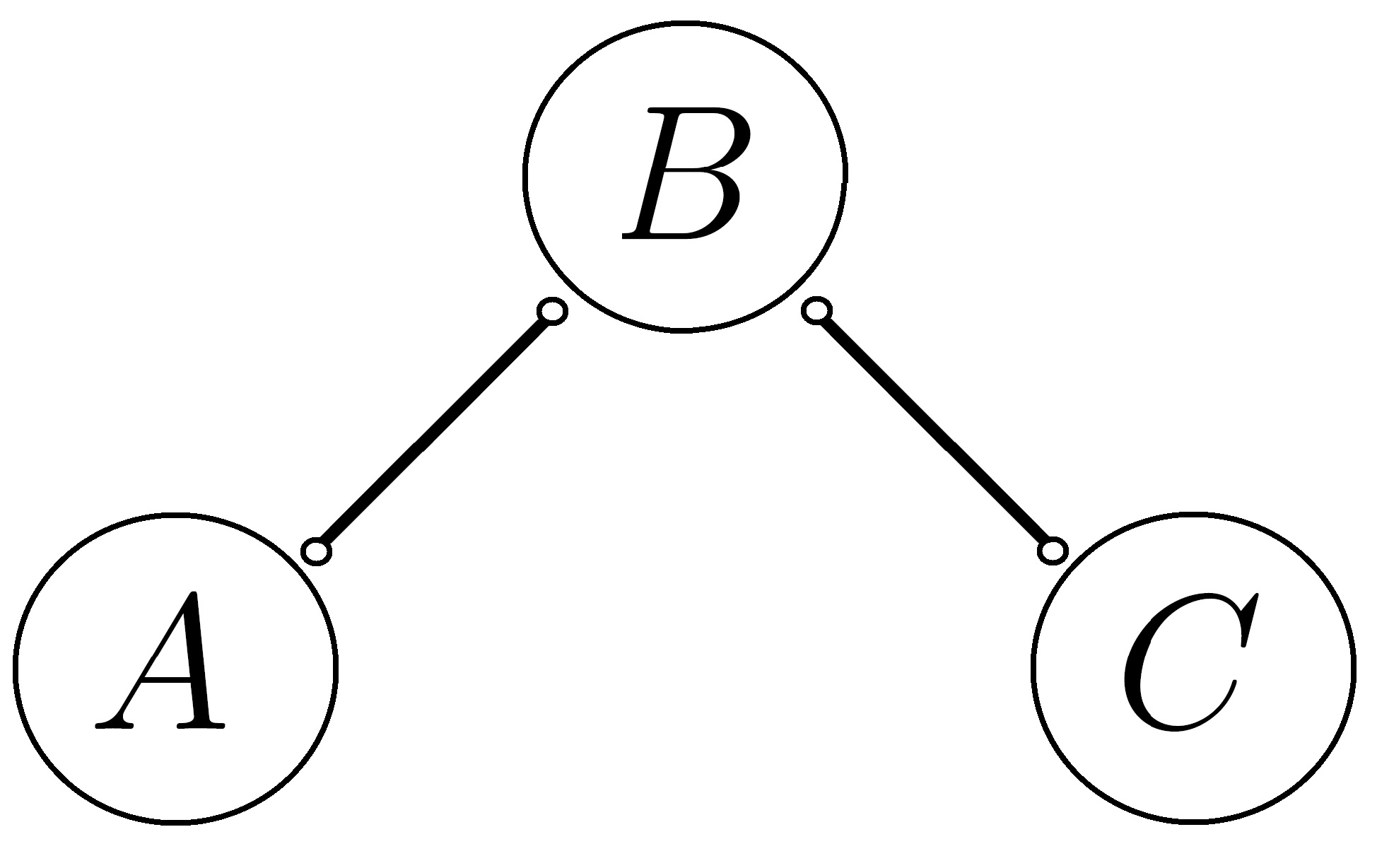}\\
  \caption{DAG representing to the output of the algorithm IC*.}
\end{figure}

\begin{figure}[h]
\centering
  \includegraphics[scale=0.08]{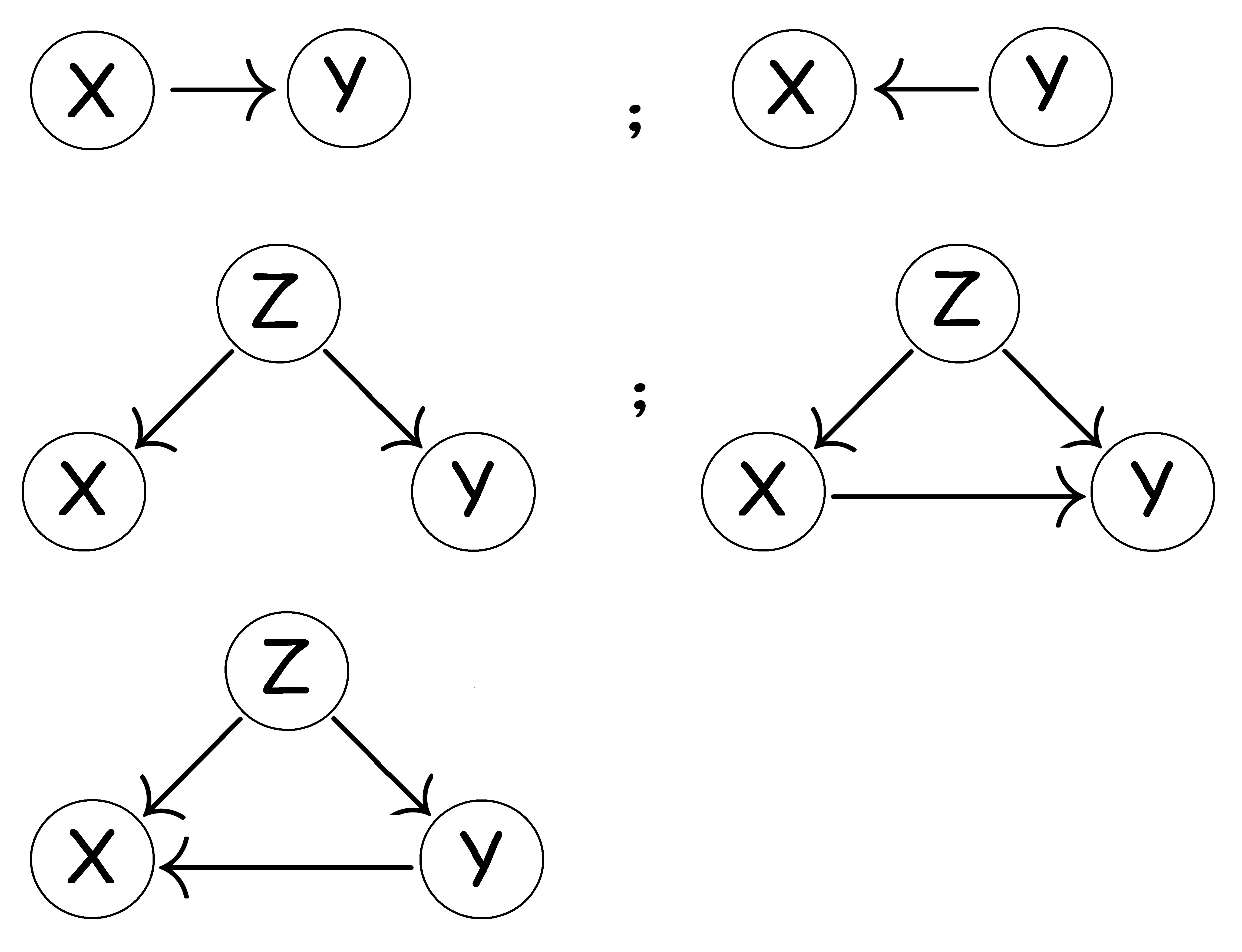}\\
  \caption{In a DAG, an undirected edge connecting the variables $X$ and $Y$ represents any of the five possibilities shown in the figure. The causal connections may be directed or mediated by a hidden variable, represented by $Z$.}
\end{figure}

The IC* algorithm eliminates combinations that create a new v-structure. The v-structure in a DAG is defined as a collision between two heads on one node, making this node a common effect of the two variables that do not exert any direct causal influence on one another. With the elimination of these configurations, the causal structures returned by the IC* algorithm are shown in Figs. 4, 5, and 6. They are separated in groups according to the ordering of the variables. The experimental proposal in \cite{Ionicioiu1} uses photons generated from parametric down-conversion. The ordering of detections can be controlled by optical delays in the photon arms.

\begin{figure}[h]
\centering
  \includegraphics[scale=0.08]{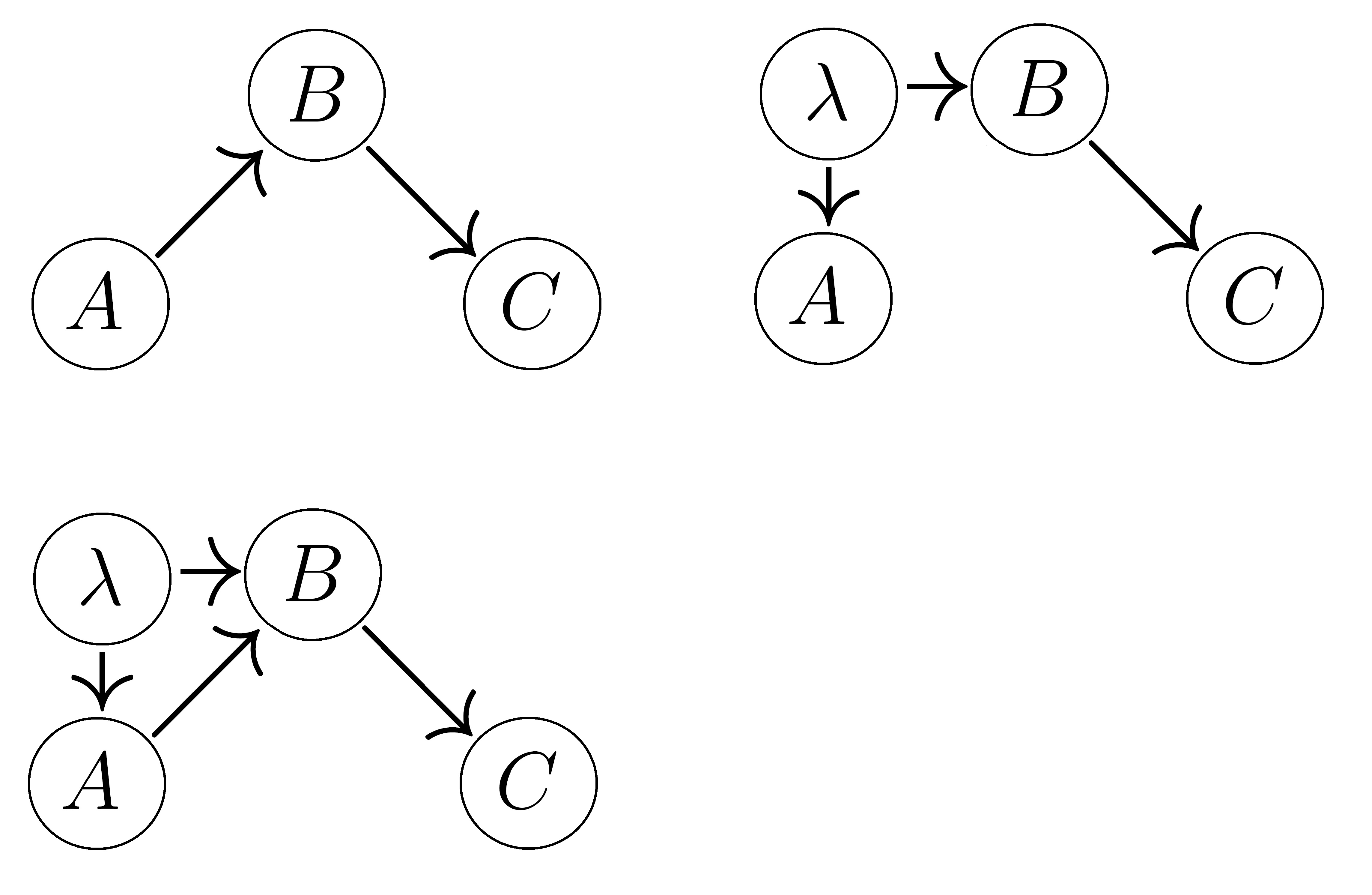}\\
  \caption{DAGs obtained when the ordering $A<B<C$ is considered.}
\end{figure}

\begin{figure}[h]
\centering
  \includegraphics[scale=0.08]{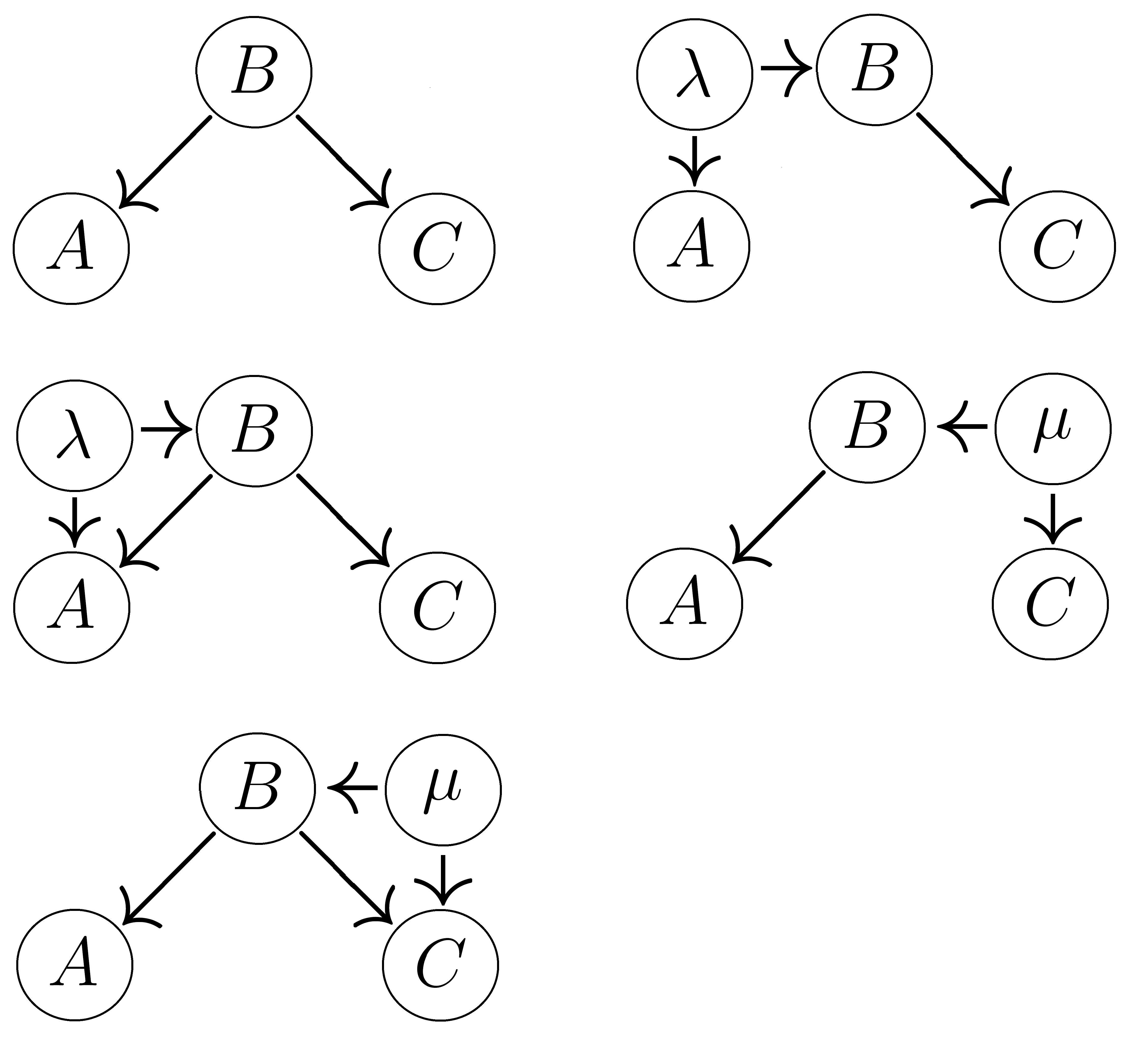}\\
  \caption{DAGs obtained when the ordering $B<A<C$ or $B<C<A$ are considered.}
\end{figure}

\begin{figure}[h]
\centering
  \includegraphics[scale=0.08]{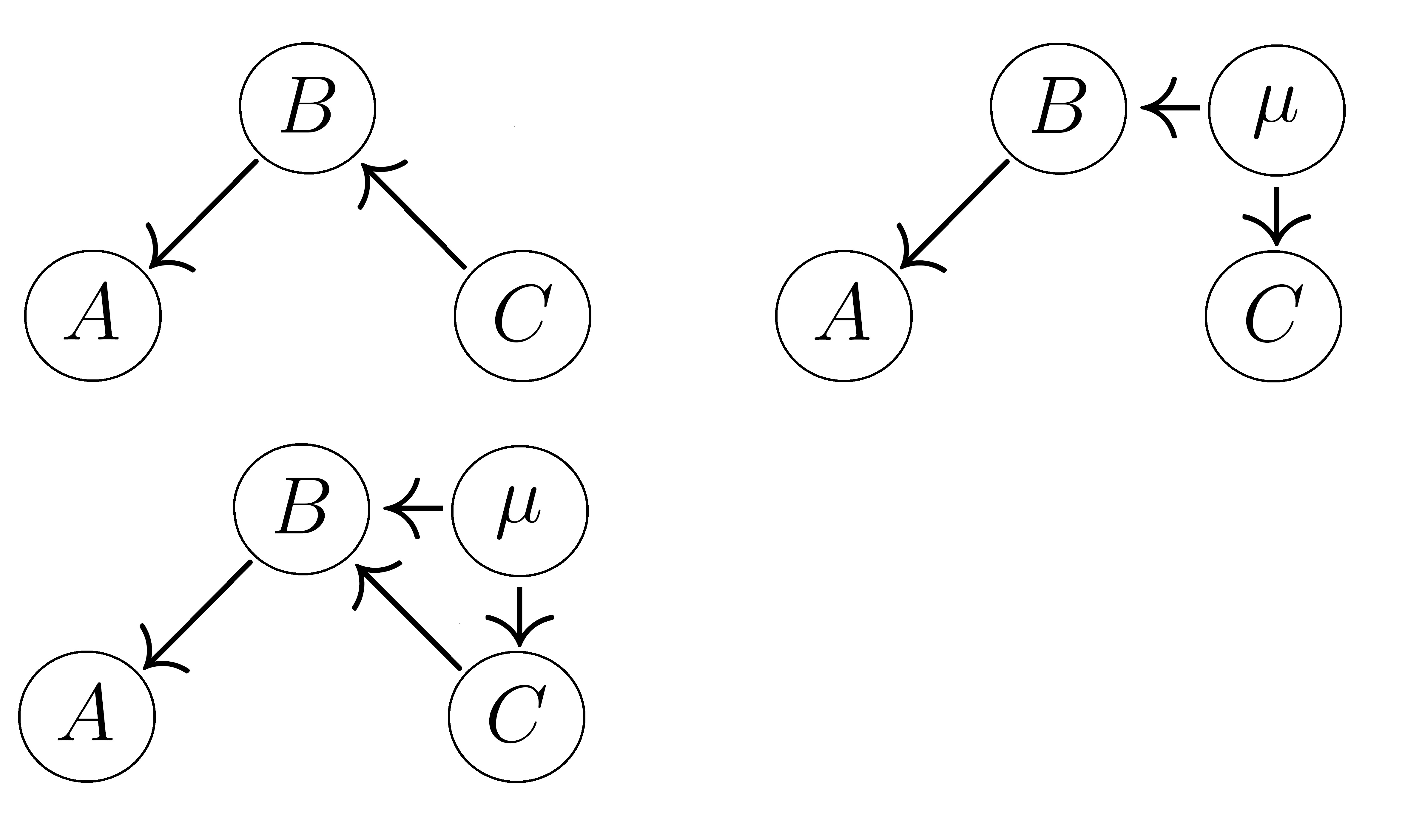}\\
  \caption{DAGs obtained when the ordering $C<B<A$ is considered.}
\end{figure}

\textit{The Results}

As IC* algorithm selects causal structures that do not create new v-structures, it is not possible to have causal structures with two hidden variables for this system. If two hidden variables ($\lambda$ and $\mu$) were considered, the first ($\lambda$), a common cause of the variables $A$ and $B$ and the second ($\mu$) , a common cause of $B$ and $C$, the variable $B$ would be a common effect of $\lambda$ and $\mu$. Therefore, conditioning on $B$ would induce a dependence between $\lambda$ and $\mu$, as these hidden variables are correlated with $A$ and $C$ respectively, it would also induce a dependence between $A$ and $C$ conditioned on $B$, contradicting the relation $A\perp\!\!\!\perp C | B$.

In the interferometric system of reference \cite{Ionicioiu1}, the space-like separation between the subsystem $q_{a}-q_{b}$ and the qubit $q_{c}$ is considered. Therefore, to avoid superluminal causation, it is necessary to assume the hidden variable $\mu$ in the causal structure to describe the correlation between $B$ and $C$. On the other hand, an objective view of the wave and particle properties that require each photon $A$ to have the intrinsic characteristic of being a particle or a wave also demands the existence of a hidden variable $\lambda$. In order for the photons $q_{a}$ to have unchanged properties (during their lifetime), the correlation between $A$ and $B$ cannot be considered as being directly caused by $B$, for it would mean that detections of $q_{b}$ would define the nature of the photons $q_{a}$. This interdiction of causal link $B \longrightarrow A$ was already present in the Classic Delayed Choice experiment, where the definition of whether or not to insert the second beamsplitter was taken after the photon enters the interferometer (the variable $B$ was determined by a QRNG in the classic setup). This directed causal connection would contradict the objective view of the wave and particle properties of the photon. Therefore, to maintain this objective view, one must consider a second hidden variable $\lambda$. In conclusion, to avoid superluminal causation and to assume the objective view of the wave and particle properties, one must consider two hidden variables. The first one ($\mu$) as a common cause of $B$ and $C$ and the second one ($\lambda$) as a common cause of $A$ and $B$. However, the application of IC* algorithm in the present system does not give causal models with two hidden variables as an output. Therefore, a conclusion from the IC* algorithm is that it is not possible to construct a causal model for this system that avoids superluminal causation and assumes an objective view of the wave and particle properties simultaneously.

For the ordering $C<B<A$, there is no causal structure with the hidden variable $\lambda$ (the common cause of $A$ and $B$). In this ordering, $A$ is directly caused by $B$, and there is no possibility to consider an objective view of the wave and particle properties even when superluminal causation is assumed (when $C$ is a direct cause of $B$). This result is also a consequence of the limitation impose by IC* algorithm regarding causal structure that creates new v-structures. If $\lambda$ was included, $B$ would be a common effect of $\lambda$ and $C$, which are variables that do not exert any direct causal influence on one another.

In the ordering $A<B<C$, superluminal causation ($B$ is a direct cause of $C$) is assumed in all causal structures. The inclusion of $\mu$ (common cause between $B$ and $C$) would create new v-structures.

For the ordering $A<C<B$ and $C<A<B$, there are no causal structures according to the IC* algorithm. In the present system, $B$ can not be a common effect because every causal structure in which $B$ is a common effect necessarily includes a new v-structure.

To summarize, in this paper, a contribution to the debate about the role of causality in quantum mechanics is presented. A recently developed mathematical tool, namely, the causal discovery algorithm is applied in the search for causal models to describe the quantum system presented in ref.\cite{Ionicioiu1}. As an output, the algorithm gives causal structures for each ordering of the detections of the variables. It is shown that no causal structure with two hidden variables is allowed for this quantum system, and a consequence of such constraint is the impossibility of constructing a causal model for this system that avoids superluminal causation and assumes an objective view of the wave and particle properties simultaneously.

\appendix*
\section{Causal Models and Causal Discovery Algorithms}

In this section a brief summary of the theory of causal models and causal discovery algorithms is given. For further details see references \cite{pearl, spirtes}.

\textit{Causal Models}

A causal model is composed of a set of statistical parameters $\Theta$ and a casual structure. The set $\Theta$ specifies the probability distribution for the variable that describe the system, and the causal structure can be represented as a directed acyclic graph (DAG) that indicates the causal connections among the variables.

Let us start by analyzing the relation between probability distribution calculation and DAGs. Consider a system composed of $n$ variables $\{X_{1},...,X_{n}\}$. The calculation of the joint probability $P(X_{1},...,X_{n})$ demands a table with $2^{n}$ entries. Using the chain rule of probability theory, one can write the joint probability as a product of each of their conditional probabilities:

\begin{equation}
P(X_{1},...,X_{n})=\prod_{j}P(X_{j}|X_{1},...,X_{j-1}),
\end{equation}       
where $\{X_{1},...,X_{j-1}\}$ are the predecessors of $X_{j}$. If the conditional probability of $X_{j}$ depends only on a subset ($Pa(X_{j})$) of its predecessors, the joint probability can be written in a simpler form given by:

\begin{equation}
P(X_{1},...,X_{n})=\prod_{j}P(X_{j}|Pa(X_{j})),
\end{equation}       
the variables in the set $Pa(X_{j})$ are called parents (or Markovian parents) of $X_{j}$.

The relations among the variables and their parents can be represented by a DAG, which is composed of directed edges connecting the nodes. In a DAG, nodes represent the variables and the directed edges correspond to connections among them. The connection between a variable and its parent is represented by an arrow form the parent to the variable. The acyclicity certifies that no paths in the graph begin and end at the same node.

In the section 1.2 of Ref.\cite{pearl}, it is shown that a necessary and sufficient condition for a probability distribution P to be represented by a DAG G is: every variable ($X_{j}$) should be independent of all its nondescendants ($Nd(X_{j})$),
conditional on its parents ($Pa(X_{j})$). This condition is called \textit{Markov Condition}, and can be written as $X_{j}\perp\!\!\!\perp Nd(X_{j}) | Pa(X_{j})$.

The interpretation of the DAGs as causal structures forms the basis of the causal model theory. In this interpretation, every parent-child connection in the DAG is a causal relation. In this context, from the analysis of a DAG, one can know the conditional independence relations associated to the probability distribution that correspond to the DAG. However, in usual scenarios of natural science, the conditional independence relations are known and the goal of the research is to obtain the causal structure that yields these relations. The tools developed in the theory of causal models that give a systematic treatment of this search are the causal discovery algorithms.

\textit{Causal Discovery Algorithms}

In this section, two principles of the causal discovery algorithms are highlighted: stability and minimality. Different causal structures can produce the same probability distribution, and consequently the same conditional independence relations. However, assuming the principles of stability and minimality, one can restrict the number of causal structures that correspond to a given probability distribution.

A causal structure is considered to be stable if its explanation for the conditional independence relations is robust to changes in the statistical parameters represented by $\Theta$. Stability is essential to assure that the probability distribution is defined by a causal structure and not by a specific configuration of parameters (fine-tuning). Therefore, the causal structures selected by the causal discovery algorithms exclude models based on fine-tuning.

The assumption of minimality results in the selection of the most simple causal structure that explains the conditional independence relations. To understand the criteria of simplicity, consider two causal models $M_{1}$ and $M_{2}$ associated with the same set of variables. If one can always find a set of statistical parameters $\theta_{2}$ that makes $M_{2}$ reproduce every probability distribution of the observed variables provided by $M_{1}$, and the contrary is not true, we say that $M_{1}$ is simpler than $M_{2}$. A simpler model is more falsifiable, therefore, it can be submitted to greater number of experimental tests. Causal models that are more falsifiable and succeed in all experimental tests are more trustable than less falsifiable models. Hence, accordingly, the causal discovery algorithms select the most simple causal structures.

As a consequence of these assumptions, the algorithm gives the output shown in Fig.2, with no connections between the variables $A$ and $C$ (that would be only possible with fine-tuning) and with the 5 possible connections (shown in Fig.3) between $A-B$ and $B-C$.

\end{document}